\documentclass{article}

\IfFileExists{neurips_2026.sty}{%
  \usepackage[dblblindworkshop, final]{neurips_2026}
}{%
  \usepackage[letterpaper,textwidth=5.5in,textheight=9in,centering]{geometry}
  \usepackage{newtxtext}
  \usepackage{natbib}
  \newcommand{\workshoptitle}[1]{}
}

\usepackage[utf8]{inputenc}
\usepackage[T1]{fontenc}
\usepackage{xcolor}
\usepackage{hyperref}
\usepackage{url}
\usepackage{booktabs}
\usepackage{tabularx}
\usepackage{microtype}
\usepackage{graphicx}

\hypersetup{hidelinks}

\definecolor{operatorpurple}{HTML}{4B1F6F}
\newcommand{\operatorquote}[1]{%
  \textcolor{operatorpurple}{\textit{``#1''}}%
}

\title{A Case Study in Assuring AI-Written Software}

\workshoptitle{Managing Agents that Manage Agents: Workshop on Responsible Use of Meta-Agents}

\author{%
Lindsey Ferris \\
Independent Researcher \\
\And
Sierra Bonilla \\
University College London \\
London, United Kingdom \\
\texttt{sierra.bonilla.21@ucl.ac.uk} \\
}

\begin{document}
\maketitle

\begin{abstract}
Software-engineering agents can enable people without formal software training to build systems they could not otherwise implement and simultaneously can produce more code than even experts can meaningfully inspect. In both cases, exhaustive code review is not reliable as the sole basis for human control. We report a case study of a production healthcare platform built through coding agents and governed by an operator without formal software-engineering training. Over time, its workflow grew into a human-led meta-agent system where one agent wrote code, other agents supervised and reviewed it, and project rules carried lessons forward. The operator found that tests, monitors and reviewing agents used to supervise the system were fallible. Some monitors measured proxies rather than outcomes, some audits failed silently, missing checks disappeared from reported results and one automated repair caused operational disruption. In this case, human control depended on keeping the intended outcome, the evidence used to judge it, the agents' permissions and the final human decision were all tied to the same underlying objective.
\end{abstract}

\section{Introduction}
Despite the name, a coding ‘agent’ is really a loop of interacting processes in which a model proposes an action, a harness executes it through tools and returns an observation, and the model iterates until the task is submitted or terminated \citep{yang2024sweagent}. Its reliability ultimately depends on whether the generated code produces the outcome the user wanted. But when the request is underspecified or the outcome is difficult to measure (i.e. ``a pretty website''), the model cannot directly optimize for it. In practice, direct line by line code review of the model output is often treated as the main form of human oversight. However, recent work suggests that experienced developers instead control coding agents through prompting, planning approval, and validation \citep{huang2025control}. One participant in a practitioner study reported that a week's commits had grown from a few hundred lines before coding agents to 10--20K after, while another described pull-request backlogs because generation outpaced review \citep{lyu2026practitioners}. Another study found that coding agents improved initial task completion but reduced code comprehension, and their initial advantage largely disappeared on a later extension task completed without the coding agent \citep{balepur2026impaired}. These results do not make source review obsolete, but they motivate complementary controls when exhaustive inspection cannot keep pace. 

There is a useful historical precedent. FORTRAN replaced large amounts of hand-written machine code with automatic translation and widened access to programming \citep{backus1957fortran}. As checking every generated machine instruction became less practical, assurance shifted toward correctness proofs \citep{mccarthy1967compiler}, conformance test suites \citep{holberton1973fortran}, per-translation validation \citep{pnueli1998translation}, verified compilers \citep{leroy2009compiler} and randomized differential testing \citep{yang2011compiler}. Coding agents are not compilers: they infer underspecified intent, act in stateful environments and may modify their own tests. They are also often proprietary and difficult to inspect directly. The analogy is limited, but the precedent is still useful: when direct inspection becomes impractical, assurance may need to move elsewhere.

We examine this question through a boundary case in which an operator without formal SE training used a coding agent as the primary developer of a production healthcare platform that she could not have implemented alone. Because exhaustive source-code review was not a realistic form of oversight, the case provides an opportunity to study what other forms of control emerged in practice. The supervisory workflow was not designed in advance. It developed through trial and error as failures revealed weaknesses and prompted new checks and review practices. Recent research provides the language of evaluation-driven development for describing parts of this evolving workflow \citep{xia2025eddops,lyu2026practitioners}. 

Using two structured interviews with the operator together with project records, we examine two questions: \textbf{RQ1: Where did human control reside when implementation and review were delegated to agents?} \textbf{RQ2: What did the resulting checks and reviewing agents contribute, and where did they fail?}

\section{Case and workflow}

\begin{figure}[t]
\centering
\includegraphics[width=0.8\linewidth]{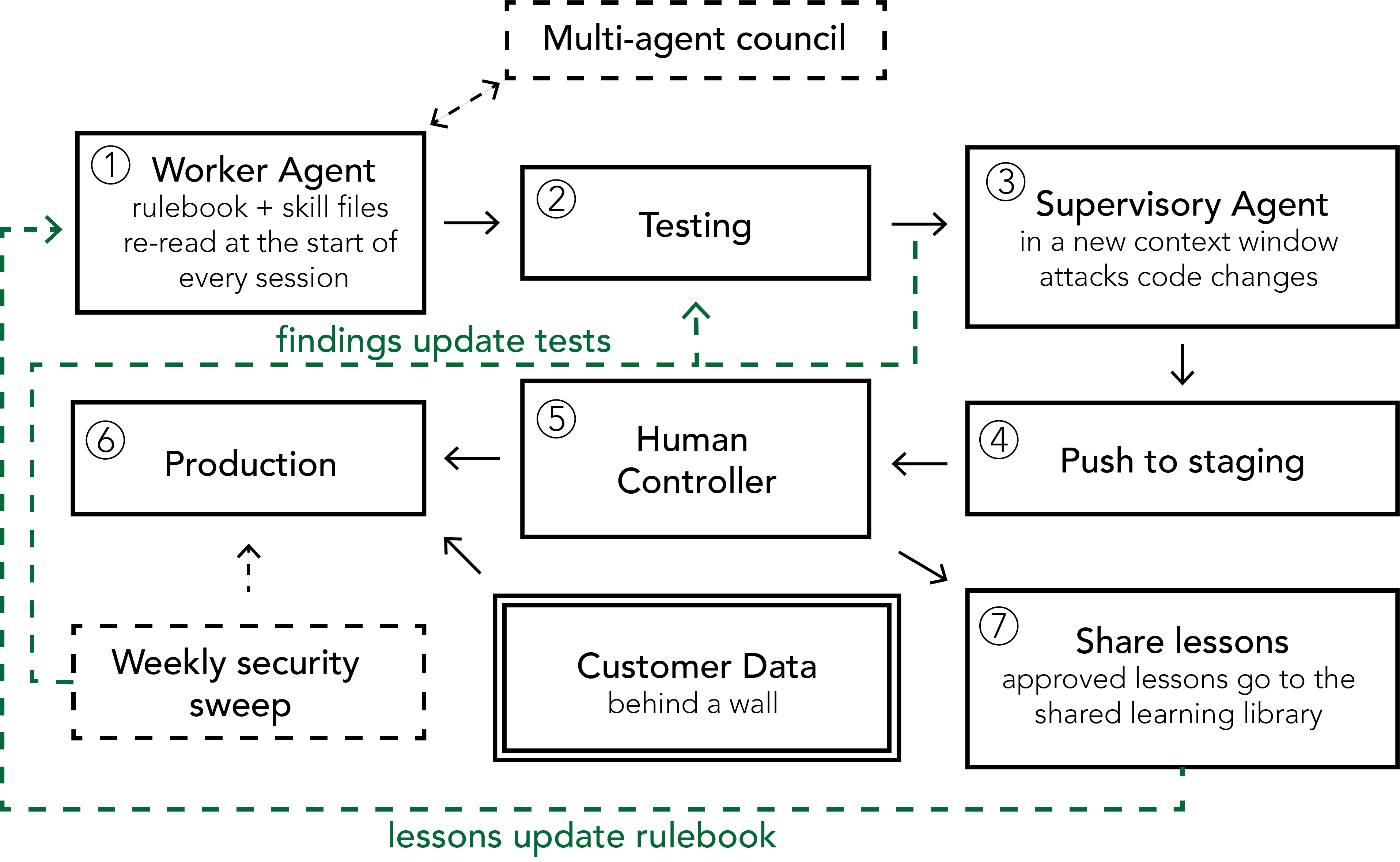}
\caption{Human-mediated meta-agent workflow used in the case study. The dashed teal arrows are the looping information where lessons update the rulebook and confirmed findings become tests.}
\label{fig:workflow}
\end{figure}

This retrospective case study draws on two structured interviews with the operator, conducted in June and August 2026, together with project records spanning six months. The records included decision logs, audit reports, incident records, check inventories and working notes. We reconstructed the development and assurance workflow from these sources and identified incidents in which an assurance mechanism produced incomplete or misleading evidence. Interview claims were cross-checked against available project records where possible.

The platform supported the business operations of a healthcare
practice, including note-taking support, scheduling and payroll workflows, alongside background functions such as access control, backups and audits. Some parts were easy for the operator to inspect: she could try a page or workflow in staging and judge whether it behaved as intended. Other properties were largely hidden from her. She could not directly see whether a backup was usable or whether an audit had run completely. Requirements of a regulated healthcare setting (US health-privacy law) applied, although we do not assess compliance. For context, the HIPAA Security Rule sets administrative, physical and technical safeguards for electronic protected health information, and NIST SP 800-66 provides implementation guidance \citep{hhs2026security,nist2024sp80066}. Coding agents could access repositories, test data and staging systems, but production data was kept outside their reach through the system architecture and file and folder permissions. 

Figure~\ref{fig:workflow} shows the full workflow. At the start of each session, the primary coding agent re-read a versioned project rulebook, current task and issue lists and reusable skill files, allowing project context and recurring procedures to carry across otherwise fresh sessions. As she explained, \operatorquote{That's how the lessons and unfinished work stick even though each session starts fresh.} The operator described the feature she wanted and approved or revised the proposed approach before implementation. Changes then passed project tests, were reviewed by a separate agent in a fresh context and were deployed to private staging for human inspection. The separate reviewer functioned as an LLM critic \citep{mcaleese2024critics}. The operator could also commission broader fresh context meta-audits that were not tied to a particular code change. These audits used newly instantiated agents without the primary development conversation to inspect wider parts of the system and its records for problems missed by the normal development loop. Their findings entered the same process of remediation, regression testing and reusable lessons.

For difficult decisions, the operator could invoke a multi-model council. Models from several providers independently answered the same brief without seeing one another's responses and returned their recommendations, reasons and risks for the operator to consider. Confirmed failures were used to search for similar problems elsewhere in the codebase and, where appropriate, to add regression tests. Agents could also propose reusable lessons at the end of a session; approved lessons were stored for use in later sessions. Production data was inaccessible to the coding agents, and changes reached production only after staging and explicit operator approval.

\section{RQ1: Where did human control reside when implementation and review were delegated to agents?}

The clearest pattern was asymmetric delegation. The operator delegated implementation and much of its technical review, but not the authority to make consequential changes. Worker and reviewing agents could propose code, findings and persistent lessons, while the operator controlled what became durable project context and what entered production; production data was also kept outside agent reach. Control was concentrated at the points where agent outputs became persistent or operationally consequential, rather than exercised continuously through line-level inspection. This resembles Barnes et al.'s framing of agentic Continuous Integration and Continuous Deployment (CI/CD) as an authority-transfer problem, with control-plane authority separated from bounded data-plane work \citep{barnes2026agency}. It also aligns with architectural tactics such as checkpoints, tool fencing and write staging for high-consequence actions \citep{safin2026autonomy}.

This also changed what the operator needed to understand. She did not need to reconstruct every implementation decision, but she did need to know what each reviewing mechanism had observed, what claim its output could support and what action could follow from it. In this case, adding another reviewer moved rather than removed part of the oversight problem: the operator also had to judge the reviewer’s evidential scope and authority. RQ2 examines what happened when those supervisory signals were themselves unreliable.

\section{RQ2: What did the resulting checks and reviewing agents contribute, and where did they fail?}
When bugs or failures were discovered, they were used to improve later development. When a problem was confirmed, the agent was asked to fix it, look for similar problems elsewhere and record any lesson that should carry forward. The operator became particularly concerned with whether these tests could genuinely detect the failures they claimed to cover: \operatorquote{I don't trust a test just because it passes.} Tests were checked by deliberately recreating targeted failures and confirming that the corresponding checks failed. This uses the broader fault-seeding intuition behind mutation testing \citep{jia2011mutation}. This is important for LLM-generated tests, because incorrect code in the prompt can substantially reduce test accuracy and bug-detection effectiveness \citep{huang2024incorrecttests}. Over time, failures changed not only the code but also the tests and instructions used in later work, similar to evaluation-driven development \citep{xia2025eddops,lyu2026practitioners}.

We also asked whether the checks accumulated through incidents and adversarial reviews covered security concerns recognized independently outside the project. The inventory was mapped against four external references that had not been consulted during its construction: the OWASP Top 10, OWASP Top 10 API Security Risks, CWE Top 25 and relevant HIPAA Security Rule provisions \citep{owasp2025top10,owasp2023api,cwe2025top25,ecfr2026hipaa}. Of 84 mapped items, 69 (82.1\%) were covered by the pre-existing checks. This suggests that the system she developed had accumulated checks addressing many concerns represented in established security guidance. It does not establish full platform security. It also does not demonstrate model generalization: because the standards are public, independence from model training cannot be assumed, a concern in benchmark-contamination research \citep{zhang2024pacost}.

However, the main finding was that these evaluations themselves became another fallible software layer. Several backup monitors checked whether intermediate events occurred without establishing that a usable backup actually existed. An audit stopped running without producing a failure, while another report calculated its pass rate only from checks that emitted results, allowing missing checks to disappear from the total. In the most consequential incident, a monitor incorrectly classified legitimate long-running work as a failure and repeatedly attempted an automated repair, causing operational disruption.  These incidents are related to the test-oracle problem; even when a check executes, its verdict may not establish the intended outcome \citep{barr2015oracle}. In each case, a reassuring signal supported a stronger claim than the evidence justified.

These incidents highlighted four recurring concerns in this case: whether a check observed the intended outcome rather than a convenient proxy; whether there was evidence that it had actually run; whether missing or empty results remained visible; and whether its authority was limited enough to prevent an incorrect result from immediately triggering harmful action. The checks and reviewing mechanisms supported later development, but their presence or apparent success was not itself sufficient evidence of reliability.

\section{Discussion and limitations}

This case is consistent with the broader finding that coding agents reorganize SE work rather than remove it. As implementation becomes easier, more human effort moves toward defining requirements, checking results and deciding whether a change is safe to release \citep{lyu2026practitioners}. The case adds a distinction between outcomes the operator could inspect directly and those accessible only through software-generated evidence. Staging worked well for observable behavior, while the clearest assurance failures occurred in less visible properties such as backups, audit execution, monitoring completeness and automated recovery.

Researchers have proposed guard agents that check other agents \citep{xiang2025guardagent}, critic models that help humans identify mistakes \citep{mcaleese2024critics}, and execution systems that preserve structured, reversible traces for meta-agent inspection and replay \citep{yu2026shepherd}. The incidents here raise a complementary issue for meta-agent research: once supervision itself is delegated, human oversight also has to account for what the supervisor observed, what its output actually supports and what authority follows from it. In this case, additional reviewers and monitors were useful, but their outputs could not be treated as ground truth.

\paragraph{Limitations and responsible use.}
This is a single case involving one operator, one platform and an evolving workflow. The operator was directly involved in creating both the system and its governance process, which may have led to an overly favourable interpretation. We reduced this risk by comparing the interview account with project records and focusing on concrete failures, but the evidence remains observational. The study does not establish that the workflow caused fewer defects, that the platform was secure or compliant, or that the same approach would be sufficient for other users or systems.

\section{Conclusion}

This case illustrates one way human control was maintained when implementation and much of its review were delegated to agents. In this workflow, control became concentrated around intended outcomes, evidence, access boundaries and consequential decisions. The same workflow also exposed a second-order assurance problem. The tests and reviewing agents introduced to support oversight could provide incomplete or misleading evidence. In this case, adding supervisory mechanisms moved part of the oversight problem to judging what those mechanisms had actually observed, what their outputs supported and what authority followed from them.

\bibliographystyle{plainnat}
\bibliography{beyond_source_code_review_references}

\appendix
\section{Implementation Details}

\paragraph{Models.}
Building, pre-merge review and meta-audits used Anthropic Claude models invoked through the Claude Code command-line agent. The multi-model council additionally queried Google Gemini and OpenAI GPT models through their APIs. Reviews and meta-audits were run by separately started agent instances with no access to the development conversation. The reviewing agents shared the builder's model family, so shared blind spots are possible.

\end{document}